\documentclass[12pt,preprint]{aastex}

\slugcomment{Submitted to Astrophysical J.}

\shorttitle{Star Formation History in Cluster Galaxies at $z\sim 0.3$}
\shortauthors{Merluzzi et al.}

\begin{document}

%% LaTeX will automatically break titles if they run longer than
%% one line. However, you may use \\ to force a line break if
%% you desire.

\title{Age, Metallicity and Star Formation History of Cluster Galaxies 
at $z\sim 0.3$\footnote{Based on observations collected at 
European Southern Observatory (ESO n. 62.O-0369, 63.O-0257, 64.O-0236)}}

%% Use \author, \affil, and the \and command to format
%% author and affiliation information.
%% Note that \email has replaced the old \authoremail command
%% from AASTeX v4.0. You can use \email to mark an email address
%% anywhere in the paper, not just in the front matter.
%% As in the title, you can use \\ to force line breaks.

\author{P. Merluzzi}
\affil{INAF Osservatorio Astronomico di Capodimonte, Napoli, Italy}

\email{merluzzi@na.astro.it}

\author{F. La Barbera}
\affil{Physics Department, Universit\`a Federico II,
    Napoli, Italy}

\email{labarber@na.astro.it}

\author{M. Massarotti and G. Busarello}
\affil{INAF Osservatorio Astronomico di Capodimonte, Napoli, Italy}

\email{michele@na.astro.it}
\email{gianni@na.astro.it}

\and

\author{M. Capaccioli}
\affil{Physics Department, Universit\`a Federico II, Napoli, Italy}
\affil{INAF Osservatorio Astronomico di Capodimonte, Napoli, Italy}
\email{capaccioli@na.astro.it}

%% Mark off your abstract in the ``abstract'' environment. In the manuscript
%% style, abstract will output a Received/Accepted line after the
%% title and affiliation information. No date will appear since the author
%% does not have this information. The dates will be filled in by the
%% editorial office after submission.

\begin{abstract}

We investigate the color-magnitude distribution in the rich cluster
AC\,118 at $z=0.31$. The sample is selected by the photometric
redshift technique, allowing to study a wide range of properties of
stellar populations, and is complete in the K-band, allowing to
study these properties up to a given galaxy mass. We use galaxy templates
based on population synthesis models to translate the physical
properties of the stellar populations - formation epoch,
time-scale of star formation, and metallicity - into
observed magnitudes and colors.  The distributions of galaxies in
color-magnitude space thus map into distributions in the space of
physical parameters. This is achieved by means of a statistical
procedure which constrains the photometric properties of AC\,118
galaxies to reproduce those of a nearby rich
cluster once evolved at $z\sim0$.  
In this way we show that a sharp luminosity-metallicity
relation is inferred without any assumption on the galaxy formation
scenario (either monolithic or hierarchical).  Our data exclude
significant differences in star formation histories along the
color-magnitude relation, and therefore confirm a pure metallicity
interpretation for its origin, with an early ($z\sim5$) formation
epoch for the bulk of stellar populations. The dispersion in the
color-magnitude diagram implies that fainter galaxies in our sample
(K$\sim18$) ceased to form stars as late as $z\sim 0.5$, in agreement
with the picture that these galaxies were recently accreted into the
cluster environment. The trend with redshift of the total stellar mass
shows that half of the luminous mass in AC\,118 was already formed at
$z \sim 2$, but also that 20\% of the stars formed at $z<1$.

\end{abstract}

%% Keywords should appear after the \end{abstract} command. The uncommented
%% example has been keyed in ApJ style. See the instructions to authors
%% for the journal to which you are submitting your paper to determine
%% what keyword punctuation is appropriate.


%% From the front matter, we move on to the body of the paper.
%% In the first two sections, notice the use of the natbib \citep
%% and \citet commands to identify citations.  The citations are
%% tied to the reference list via symbolic KEYs. The KEY corresponds
%% to the KEY in the \bibitem in the reference list below. We have
%% chosen the first three characters of the first author's name plus
%% the last two numeral of the year of publication as our KEY for
%% each reference.

\section{Introduction}\label{INTRO}
The color-magnitude relation (CMR) of cluster early-type galaxies has
been extensively investigated at $z < 1$ to trace their star formation
history and hence to constrain their formation epoch (e.g. e.g Kodama
\& Arimoto 1997, hereafter KA97; Ellis et al. 1997; Gladders et
al. 1998; Stanford, Eisenhardt, \& Dickinson 1998; Kodama \& Bower
2001, hereafter KB01; Smail et al. 2001; van Dokkum et al. 2001).

The most important observational results are: i) the slope of the CMR
does not depend on redshift; ii) the optical-NIR rest-frame colors of
early-type cluster members become bluer with increasing redshift; iii)
the intrinsic scatter in the optical-NIR colors of early-type galaxies
is small at all redshifts (e.g. Ellis et al. 1997; Stanford et al.
1998; van Dokkum et al. 1998, 2000; Kodama et al. 2001). These points
lead to explain the color-magnitude (CM) sequence as a correlation
between galaxy mass and metallicity, while the age of galaxies play
only a marginal role, if any (e.g. KA97).  Two different scenarios can
successfully explain the CMR as function of redshift: the monolithic
collapse (e.g. Eggen, Lynden-Bell, \& Sandage 1962; Tinsley \& Gunn
1976) and the hierarchical merging (e.g. Kauffmann 1996, Kauffmann \&
Charlot 1998).  In the former, the trend of the mass-metallicity
sequence is explained by the fact that the more massive galaxies
retain supernova ejecta more effectively, resulting in higher
metallicities and hence in redder colors for more luminous galaxies
(e.g. Arimoto \& Yoshii 1987; KA97). The unchanged scatter of the
colors of early-type galaxies with redshift indicates either that the
galaxies assembled synchronously over redshifts (at least for $z<1$)
or that they stochastically formed at much earlier times (see Ellis et
al. 1997).  For what concerns the alternative picture, Kauffmann \&
Charlot (1998) claimed that the CMR can be reproduced in a
hierarchical merging picture, where the more massive/metal-rich
ellipticals result from mergers of massive/metal-rich progenitor disk
galaxies. In both scenarios the color evolution of early-type cluster
galaxies is in agreement with the passive evolution of an old stellar
population formed early in the past (see also Stanford et al. 1998;
Kodama et al. 1998).

Both the main evolutionary scenarios have to face with the
evidence for the presence of a significant population of blue galaxies
in rich cluster environments at $z\geq$ 0.2, as shown for the first time
by Butcher \& Oelmer (1978) and has confirmed by several photometric and
spectroscopic observations (e.g. Butcher \& Oelmer 1984; Couch \&
Newell 1984; Ellis et al. 1985; Dressler \& Gunn 1982; Couch \&
Sharples 1987, Couch et al. 1994; Dressler et al. 1994).  Taking into
account a representative sample of the whole cluster population, KB01
re-investigated the photometric Butcher-Oelmer (B-O) effect
in distant clusters. They found that the passive evolution of galaxy 
populations can reconcile the B-O effect with the tight CMR of the Coma 
cluster. Furthermore, KB01 found that
the distribution in the color-magnitude diagrams suggests a scenario where 
star formation of galaxies accreted by the cluster declines on a 1 Gyr 
time-scale and it is not sharply truncated by interaction with the cluster 
environment. In this scenario, the B-O effect depends on the decline of star
formation of field galaxies when they are accreted into the cluster and on 
the decline of the rate of accretion of new galaxies at lower
redshifts.  

In the present work we will apply the CM diagram to gain insight into
the star formation history in the galaxy cluster AC\,118 at $z=0.31$.
We will use population synthesis models in order to describe the
observed CM distribution of galaxies in AC\,118 in terms of stellar
populations parameters. The cluster sample is selected according to
the photometric redshift technique, and is complete in K-band,
avoiding biases introduced by measuring the blue wavelengths in the
cluster rest-frame.  The early-type galaxy population in the core of
AC\,118 was already analyzed by Stanford et al. (1998) who found
evidence in favor of the passive evolution scenario.  A spectroscopic
study of the cluster was performed by Couch \& Sharples (1987) and
Barger et al. (1996) who claimed for recent ($\lesssim 2$ Gyr) bursts
of star formation.

The layout of the paper is the following.  In \S~2 we describe the
sample of galaxies at $z\sim 0.3$.  In \S~3 we introduce the galaxy
templates that will be used to interpret the observed photometry in
terms of physical properties of stellar populations and we describe our
approach. The resulting distribution of the physical parameters is
analyzed in \S~4, where we also discuss the origins of the
CMR and the global star formation history.  In
\S~5 we summarize the main aspects of the work and draw the
conclusions.  In the following we assume $\Omega_m=0.3$,
$\Omega_\Lambda=0.7$ and $\mathrm{ H_0= 70~Km s^{-1} Mpc^{-1} }$. With
this cosmology the age of the universe is $\mathrm{13.5~Gyr}$, and the
redshift of AC\,118 corresponds to a look-back time of
$\mathrm{3.5~Gyr}$. We verified that changing the cosmology does not
affect the results of the present work.

\section{The Sample at $z\sim0.3$}

\begin{figure}
\epsscale{0.75}
\plotone{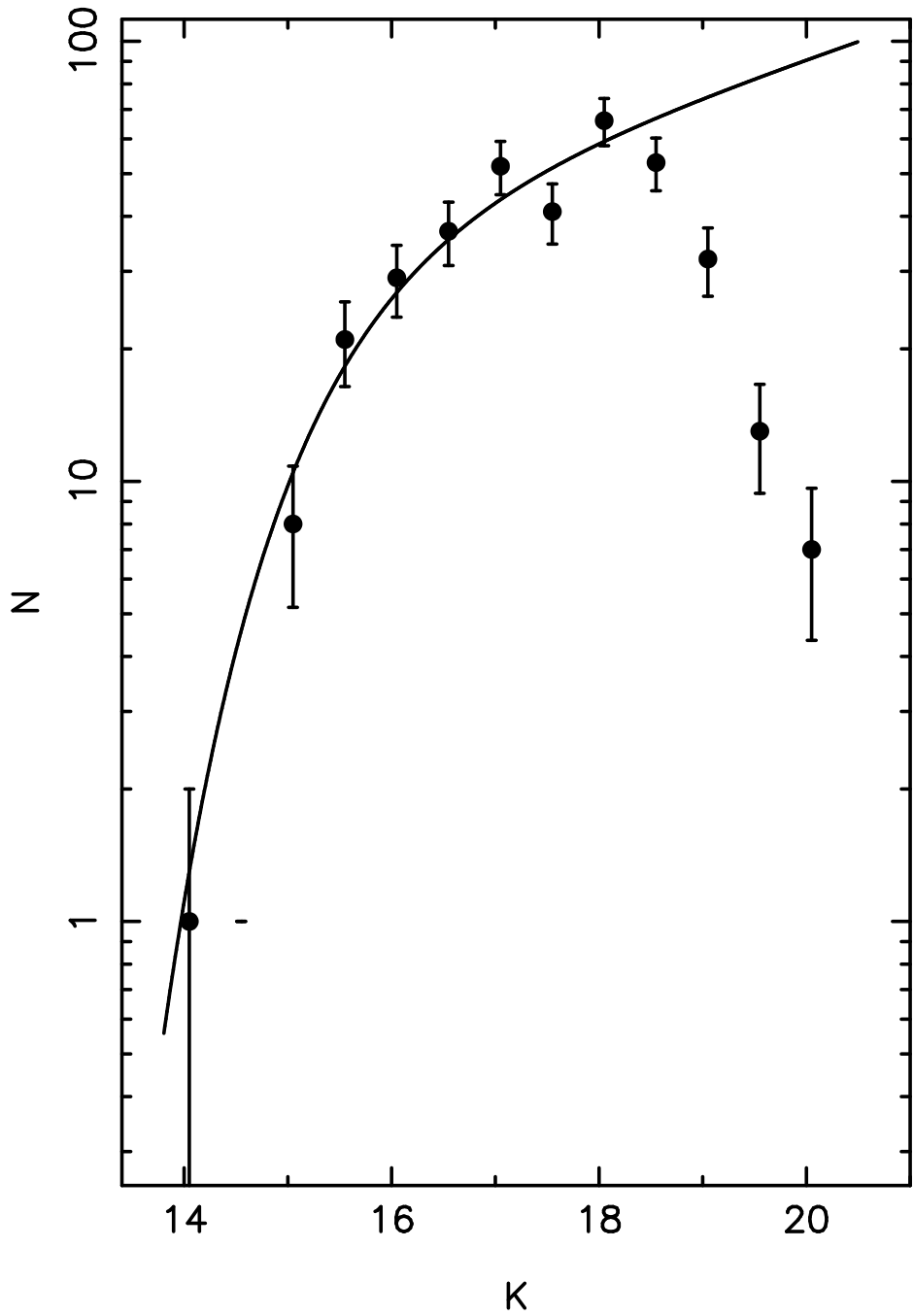}
\caption{The distribution of K-band magnitudes for the galaxies
defined as cluster members by the photometric redshift technique (dots) is 
compared with the K-band luminosity function of AC\,118 by Andreon (2001;
solid line). The completeness limit of the photometric redshift sample turns
out to be K=18.25.
\label{SELZP}}
\end{figure}

The present analysis is based on VRIK photometry for a sample of
galaxies in a field of $6.0 \times 6.0 ~ \mathrm{arcmin}^2$ ($1.6
\times 1.6 ~ \mathrm{Mpc}^2 $ at $z=0.31$) centered on the galaxy
cluster AC\,118. The optical (VRI) data are taken from the catalog in
Busarello et al. (2002), which also includes photometric redshifts,
while the K-band photometry is described in Andreon (2001).  The
present sample was selected according to the following criteria: a)
galaxies are cluster members according to their photometric redshifts
and b) the sample is complete in the K-band.  In Figure~\ref{SELZP} we
compare the distribution of the K-band magnitudes for the 459 member
galaxies from Busarello et al. (2002) with the K-band luminosity
function of AC\,118 by Andreon (2001), obtained by statistically
subtracting field counts. The Figure shows that the trend of our
counts and the luminosity function of AC\,118 are consistent down to
$\mathrm{K}=18.25$, suggesting, therefore, that the sample of member
galaxies is fairly complete down to this limit.  This leads to a
final sample of $\mathrm{N}=252$ galaxies brighter than
$\mathrm{K}=18.25$. In order to quantify the field contamination in
the redshift range adopted to select cluster members (i.e. $z \in
[0.24, 0.38]$), we note that the field population at $z\sim0.3$ is
dominated by late-type galaxies bluer than $\mathrm{I-K=2.0-2.5}$, and
therefore, the field contaminants in our sample are expected to
be brighter than $\mathrm{I\sim20.5}$. According to the Canada France
Redshift Survey (Lilly et al.~1995), we expect $\sim15$ galaxies down
to $\mathrm{I=20.5}$ in the cluster area and redshift range, amounting to $6\%$
of the galaxies in the final sample. Since this estimate is an upper limit
of the number of field contaminants, we conclude that
foreground/background contamination is not statistically relevant to
the present analysis.

The color indices were measured within a fixed circular aperture of
diameter $4.4''$ ($20 ~ \mathrm{kpc}$ at $z=0.31$). In the following,
we will also use the $\mathrm{V\!-\!K}$ colors derived by Bower,
Lucey, \& Ellis (1992a) for galaxies in the Coma cluster within an
aperture of $10''$, which corresponds to $\sim 7$ kpc.  Since galaxies
are known to have internal color gradients, a suitable comparison
between different redshifts must take into account the physical size
of the aperture within which galaxy colors are derived. However, as
shown by Kodama et al. 1998 (see their Figure~3), the correction from
the $\mathrm{\sim 7~kpc}$ aperture to the $20~\mathrm{kpc}$ aperture
turns out to be negligible for the Coma galaxies.

\begin{figure}
\epsscale{0.75}
\plotone{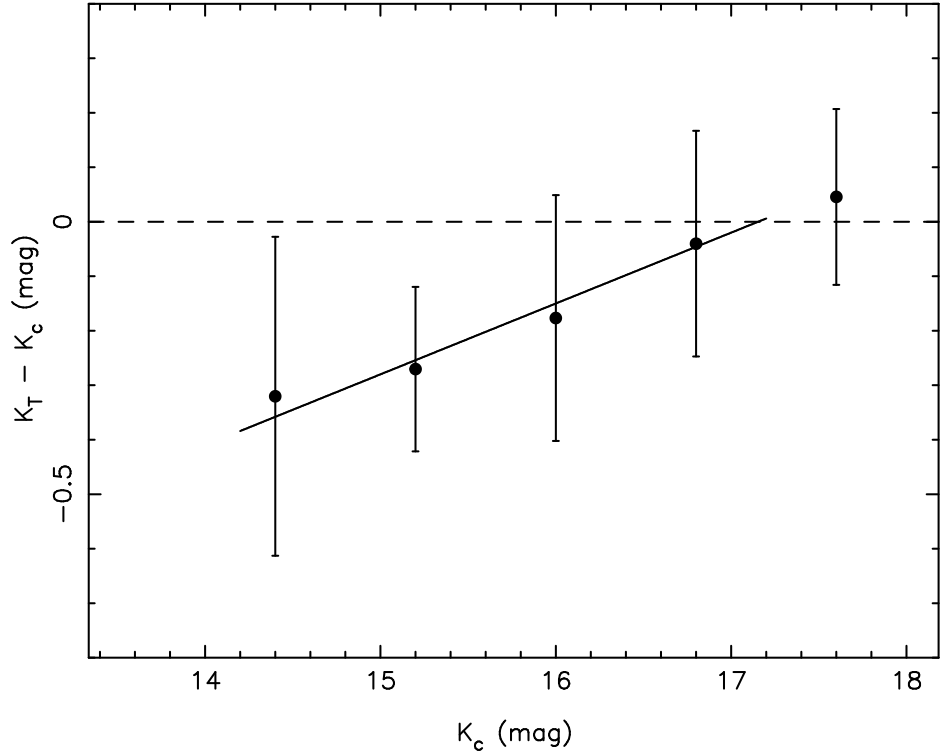}
\caption{Comparison of zeropoint-corrected Kron ($K_c$) magnitudes
with the total ($K_T$) magnitudes derived from the structural
parameters analysis, together with the best linear fit (solid line)
used to derive our total magnitudes.  The error bars show the standard
deviation of the difference between $K_T$ and $K_c$ in each bin of 0.8
mag.}
\label{KCKT}
\end{figure}

In order to estimate the total magnitude, we used adaptive apertures
of radius $\alpha \cdot r_c$, where $r_c$ is the Kron radius (see Kron
1980). We chose $\alpha=2.5$, for which the Kron magnitude is expected
to enclose $94\%$ of the total flux of the object (see Bertin $\&$
Arnouts 1996), and to correct for this factor, we added to the Kron
magnitudes the term $2.5 \log(0.94)$.  Since the bright cluster
galaxies have extended halos\footnote{High values of the Sersic index
$n$.}, the estimate of the total magnitude requires a large
extrapolation of the light profile.  To account for this
fact, it is necessary to correct the Kron magnitudes $K_c$ of the
brightest galaxies. To this aim, we compared the $K_c$ values with
those derived by the two-dimensional fit of the surface brightness
distribution for the subsample of $\mathrm{N}=95$ galaxies analyzed in
La Barbera et al. (2002). The comparison is shown in Figure~\ref{KCKT}
as a function of $K_c$.  We found that the Kron magnitude
underestimates the galaxy luminosity for values of $K_c$ brighter than
$\mathrm{K\sim17}$. The trend in Figure~\ref{KCKT} is described by the relation
$K_T-K_c = 0.13 \times K_c-2.23$ ($K_c < 17.2$), that was used to
correct the values of $K_c$ for each galaxy in our sample.

\begin{figure}
\epsscale{0.9}
\plotone{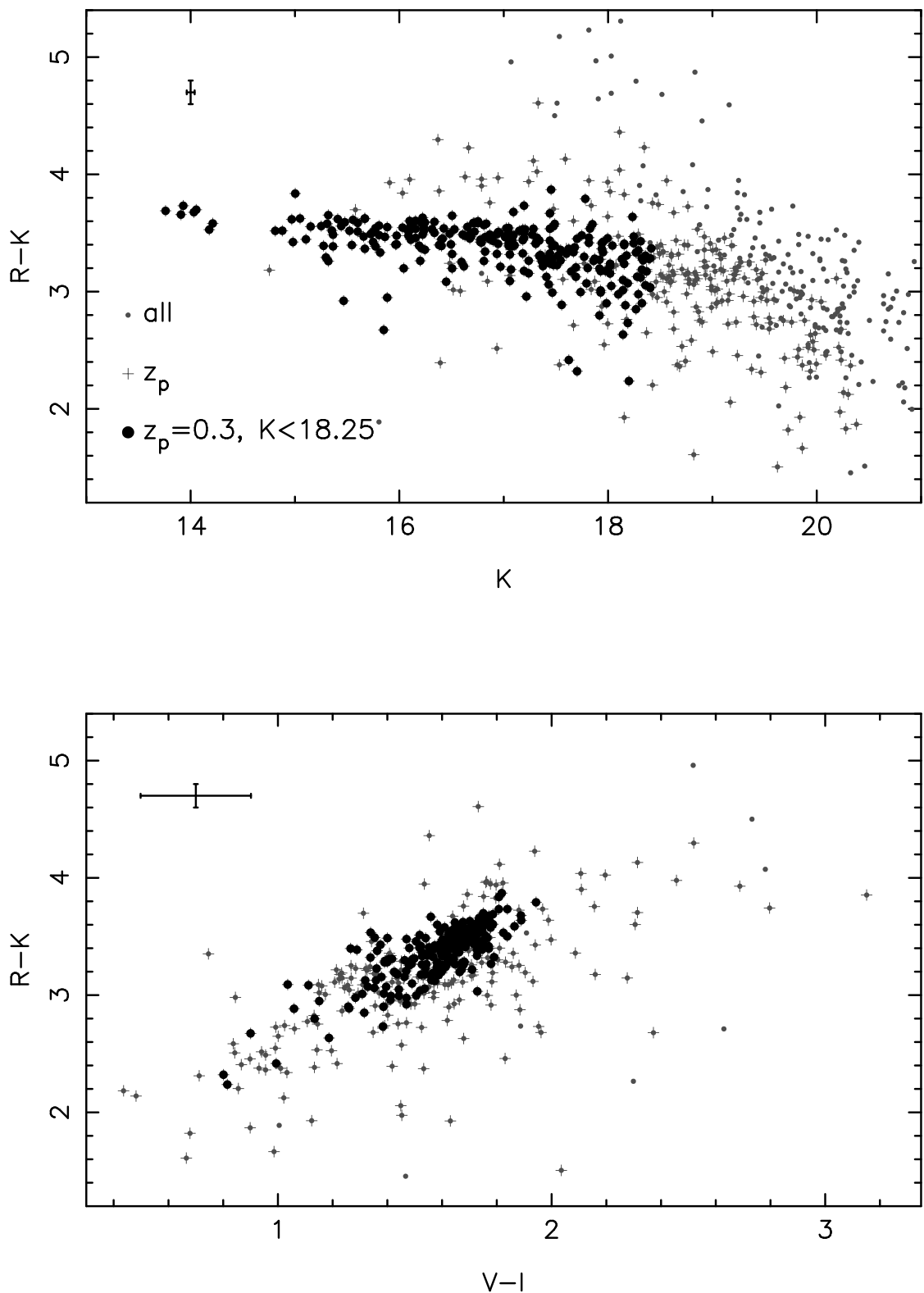}
\caption{Distribution of galaxies in the field of AC\,118 in the
($\mathrm{K,R\!-\!K}$) and ($\mathrm{V\!-\!I}$,$\mathrm{R\!-\!K}$) diagrams 
(upper and lower
panels respectively). For each plot, the typical photometric errors
(3$\sigma$ confidence intervals)
are shown in the upper-left corner. Gray dots: all the galaxies in the
K-band field; crosses: galaxies with photometric redshift; black filled 
circles: complete K-band sample of cluster members.  }
\label{CMCCTOT}
\end{figure}

In Figure~\ref{CMCCTOT} we show the CM distributions of 1) all the
galaxies in the K-band field, 2) the galaxies with available
photometric redshift and 3) the $\mathrm{N}=252$ galaxies of the
sample considered in the analysis.

\section{Modelling the Evolution of Stellar Populations }\label{MODEL}

Our goal is to fit the colors and magnitudes of galaxies in AC\,118
by imposing that their evolution at $z\sim0$ reproduces the properties
of the CM diagram of a nearby galaxy cluster. To this aim, we use
stellar population models at different evolutionary stages.

Since AC\,118 is a rich, Coma-like, cluster with high X-ray
luminosity, the properties of its galaxy population have to be
compared with those of galaxies in a rich cluster at $z\sim0$. The
sample of galaxies at $z\sim0.3$ covers the central cluster region
($6.0 \times 6.0 ~ \mathrm{arcmin}^2$), which corresponds to an area
of radius $\mathrm{\sim1~Mpc}$. In this region, rich nearby clusters
are very similar in their photometric properties. The galaxy
population is dominated by early-type galaxies which follow a tight CM
relation (see Bower et al. 1992a, Bower, Lucey, \& Ellis 1992b), with
few percents of galaxies having bluer colors (see Butcher \& Oemler
1978). Therefore, we can constrain the properties of the galaxies
using only the overall features of the CM diagrams at $z\sim0$,
without comparing the properties of AC\,118 galaxies with those of a
specific nearby cluster.

We describe the stellar populations in terms of their formation
epoch $t_0$, time scale $\tau$ of star formation and metallicity
$Z$. We do not assume any {\it a priori} probability distribution of $
\{t_0,\tau,Z \}$, but instead we derive it by comparison of a set of
model (template) galaxies with the observed CM distributions.

\subsection{Galaxy templates and the constraints at $z\sim0$}
\label{galtemp}

The galaxy templates were obtained by the GISSEL98 synthesis code of
Bruzual \& Charlot (1993). Each template is defined by the three
physical parameters $t_0$, $\tau$ and $Z$. The code allows to build
galaxy templates with metallicity in the range $Z=$0.0001 -- 0.1
\footnote{With intermediate values $Z= 0.0004, 0.004, 0.008, 0.02,
  0.05$.}  and predicts the template properties at 220 steps in age
ranging from $t=1~\mathrm{Myr}$ to $t=20 ~\mathrm{Gyr}$.  The star
formation rate is chosen in the form $e^{-t/\tau}$, with $\tau$ in the
range 0.01 -- 15.0 Gyr, with a Scalo (1986) initial mass function.  
For each value of $ \{t_0,\tau,Z \}$ in the grid of input
parameters, we computed the magnitudes in the V-, R-, I- and
K-band. For all other values of $ \{t_0,\tau,Z \}$, the
magnitudes were derived by interpolation.

Since the magnitudes of the GISSEL98 templates are arbitrarily
normalized to one solar mass, they are defined within an additive term,
and therefore they cannot be directly compared to the observed
magnitudes. We derived the additive term by using the properties of
the CM distribution at $\mathrm{z\sim0}$.  To this aim, for each
template we computed the $\mathrm{V\!-\!K}$ color at $z\sim0$ and
compared the template magnitude in the K-band with that expected for a
galaxy in a rich nearby cluster with the same $\mathrm{V\!-\!K}$
color. We used the ($\mathrm{K,V\!-\!K}$) CM distribution at $z\sim0$
because i) the CM relation in the ($\mathrm{K,V\!-\!K}$) plane has
well known properties (Bower et al. 1992a, Bower et al. 1992b) and ii)
the V- and K-band at $z\sim0$ correspond approximately to the same
rest-frame of R- and K-band\footnote{The K-band at $z\sim0.3$ matches
the H-band rest-frame. However, the $\mathrm{H\!-\!K}$ color is almost
independent of the galaxy spectral type, and therefore the difference
between the K- and the H-band rest-frame magnitudes is not relevant
for the present analysis.
\label{nota1}} at $z\sim0.3$.

In order to derive the magnitude expected for a galaxy at $z\sim0$
with a given $\mathrm{V\!-\!K}$ color, we took advantage the following
properties of galaxies in the Coma cluster.

\noindent
{\it 1--Red sequence.} To describe the ($\mathrm{K,V\!-\!K}$) red
sequence, we used of the CM relation by Bower et al. (1992a):
\begin{equation}
\mathrm{ V-K = (-0.07 \pm 0.013) \cdot K + 3.92 \pm 0.20}  \ \ .
\label{CM_COMA}
\end{equation}
The intrinsic dispersion of this relation is $\sigma_{\mathrm{V-K}}\sim0.03$
mag along the color direction.

\noindent
{\it 2--Blue galaxies.} We described the distribution of galaxies
below the red sequence by using the properties of the CM diagram for
the Coma cluster recently studied by Terlevich, Caldwell, \& Bower
(2001).  The sample is complete down to $\mathrm{K\sim16}$
($\mathrm{K\sim13}$ for E/S0 templates at $z\sim0$), that corresponds
approximately to the completeness limit of AC\,118 evolved to
$z\sim0$.  In Figure~5 of Terlevich et al. (2001) we notice that most of
the objects within the completeness limit follow a tight
$\mathrm{U\!-\!V}$ CMR, while a small fraction of Sp/Irr galaxies
($\sim 5\%$) are located in a rectangular region with
$\mathrm{14.5<V<16}$ ($\mathrm{11.5<K<13}$) and with significantly bluer
colors $\mathrm{0<U\!-\!V<0.4}$ ($\mathrm{2.6<V\!-\!K<2.9}$) with
respect to the red sequence.

\noindent
{\it 3--Luminosity function.} We adopted the luminosity function (LF)
in the K-band for the Coma cluster. To this aim, we used the H-band LF
by de Propris et al. (1998) and Andreon \& Pell\'o (2000) for the
central regions of the Coma cluster, corrected (see note~\ref{nota1}) 
by the color term $\mathrm{H\!-\!K=0.22~mag}$.

Each template describes either i) a galaxy of the red sequence, or ii)
a blue (Sp/Ir) galaxy if it lies $3\sigma_{\mathrm{V-K}}$ below the CM
sequence of Eq.~\ref{CM_COMA}.  In case i), the K magnitude is
obtained by a normal deviate of width
$\sigma_{\mathrm{V\!-\!K}}/b_{\mathrm{CM}}$, where $b_{\mathrm{CM}}$
is the slope of the ($\mathrm{K,V\!-\!K}$) CM relation (see point {\it
1}), while in case ii) the magnitude is obtained by a uniform
distribution with extremes K=11.5 and K=13 (see point {\it 2}).  In
both cases, the magnitudes were extracted by the adopted distributions
(normal or uniform), using as weighting factor the K-band LF at
$z\sim0$ (see point {\it 3}).

The magnitudes of each template at $z\sim0.3$ were corrected
by taking into account the corresponding additive terms and the
luminosity distance term relative to the redshift of AC\,118.  

\subsection{The fitting procedures}
\label{FITMOD}
For each galaxies in AC\,118, we derived the `best' values of
$\{t_0,\tau,Z \}$ by two different fitting procedures.  In case {\it a)},
we obtained $\{t_0, \tau, Z \}$ by minimizing for each galaxy the
distance of the models from the observed point in color-color space at
$z \sim 0.3$, that is by minimizing the function:
\begin{equation}
\rm \chi^2_{AC118}(t_0,\tau,Z) =
\left[(V-I)_{j}-(V-I)_{templ}\right]^2+\left[(R-K)_{j}-(R-K)_{templ}\right]^2,
\label{eqchiACa}
\end{equation}
\noindent
where the subscript $\mathrm{j}$ denotes the galaxies of the AC\,118
sample, while the subscript $\mathrm{templ}$ refers to the photometric
quantities of the templates, which are functions of $t_0$, $\tau$ and
$Z$.  In this case, the choice of the best templates depends only on
the photometric properties of galaxies at $z \sim 0.3$ without any
constraint at $z\sim0$. We point out that this procedure is completely
independent of the photometric properties of the local cluster.

In case {\it b)} the `best' values of $ \{t_0,\tau,Z \}$ were obtained by
minimizing the function:
\begin{equation}
\rm \chi^2_{AC118}(t_0,\tau,Z) =
\left[K_{j}-K_{templ}\right]^2+\left[(V-I)_{j}-(V-I)_{templ}\right]^2+
\left[(R-K)_{j}-(R-K)_{templ}\right]^2.
\label{eqchiACb}
\end{equation}
\noindent
In this case the choice of $\{ t_0,\tau,Z \}$ is also driven by the
template K-band magnitudes, which were scaled as described in the
previous section. In such a way, we are constraining the template of
each galaxy in AC\,118 to occupy the red sequence locus or the region
populated by blue galaxies (points {\it 1} and {\it 2} of
Section~\ref{galtemp}) when evolved to $z\sim0$. This constrains, 
therefore, AC\,118 to
belong to the same evolutionary sequence of a rich nearby cluster.  We
point out that this procedure does not imply that the set of N=252
best fitting templates of AC\,118 galaxies, when evolved to $z\sim0$,
reproduces a CM diagram with the same properties of that observed for
a nearby rich cluster, i.e. the slope and the intrinsic scatter of the
CM relation, the fraction of blue galaxies and the LF. In fact, our
unique constraint is that each template is bounded by the same region
of galaxies in the CM diagram at $z\sim0$. This point will be further
discussed in Section~\ref{FINE}.

To account for measurement errors, the fitting procedures
were iterated by shifting colors and magnitudes of galaxies in AC\,118
according to their photometric uncertainties\footnote{The shifts were
assigned by taking into account also the correlation between the
measurement errors on colors and magnitudes.}. In this way, for each
iteration we obtained a distribution of `best' parameters
$ \{t_0,\tau,Z \}_j \ , \ j=1...N$ which describes the
photometric properties of all the $\mathrm{N}=252$ galaxies in our
sample at $z\sim 0.3$. Since the distributions of $\{t_0,\tau,Z \}$
coming from the different iterations are practically identical, in the
following we will discuss the results by averaging the properties of the
different distributions of `best' parameters.

\section{Ages, Star Formation Rates and Metallicities}\label{RESULTS}

\begin{figure}
\epsscale{1.05}
\plotone{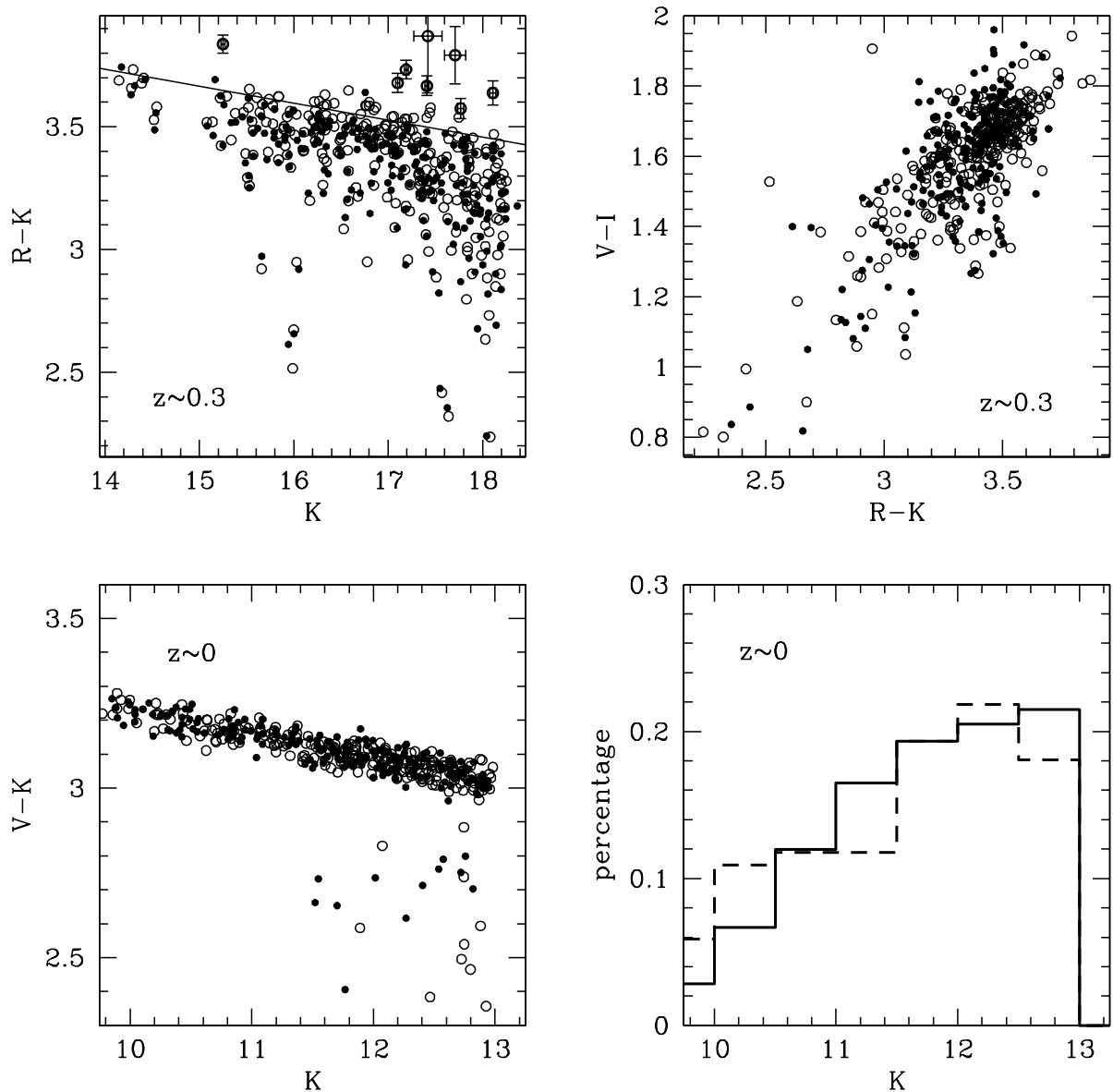}
\caption{{\it Upper panels}: distributions in colors and
magnitudes of the best model (filled circles) and of the sample at
$z\sim0.3$ (empty circles).  {\it Left bottom panel}: the CM
distribution at $z\sim0$. {\it Lower right panel}: luminosity
distribution at $z\sim0$ of the model (dotted line) and of the local
sample (continuous line).  }
\label{MODOBS}
\end{figure}

In Figure~\ref{MODOBS} we compare the distributions in the CM space of
the best fitting templates obtained according to case {\it b)} of
Section.~\ref{FITMOD} with those of the sample at $z\sim0.3$ and with
the CM distributions expected for a nearby rich cluster.  This local
sample is obtained by using the same recipe used for deriving the
additive terms of galaxy templates in Section.~\ref{galtemp}.  First,
we generated a set of magnitudes according to the Coma K-band LF (see
point {\it 3} of Section.~\ref{galtemp}). Then we assigned to each magnitude a
$\mathrm{V\!-\!K}$ color according to the CMR of the Coma cluster (see
point {\it 1} of Section.~\ref{galtemp}) and by imposing that the number of
blue galaxies in the CM plane amounts to $5\%$ (see point {\it 2} of
Section.~\ref{galtemp}).

In Figure~\ref{MODOBS}, the distributions of the model match those
observed for AC\,118, with the exception of few points whose
$\mathrm{R\!-\!K}$ colors are too red with respect to the
templates. In order to address this problem, we introduced a red
envelope of the CMR of AC\,118, defined as the locus in the plane
($\mathrm{K,R\!-\!K}$) corresponding to the reddest stellar
populations among the considered templates (cfr.~KB01). To this aim,
we considered simple stellar populations with formation epoch equal to
the age of the universe and different metallicities.  It turns out
that $\sim15\%$ of the galaxies in AC\,118 are located above the red
envelope. Six objects deviate by more than $0.1~\mathrm{mag}$, while
the photometric errors are not large enough to explain this
difference. We will come back to this point at the end of Section~4.4.

When evolved to z$\sim$0, the CM distribution of the best models for
AC\,118 gives a reliable description of the CM distribution of the
local simulated sample: most of galaxies follow a tight CMR with slope
and scatter consistent with those of the CMR of Coma, while few
galaxies (4\%) lie in the blue-faint area of the CM diagram. It is
worth to be noticed that the derived luminosity function also matches
that of the Coma cluster.

\subsection{Distributions in the Parameter $\{t_0,\tau,Z \}$ Space}

In order to analyze the allowed ranges of physical parameters, 
we compare the distributions of $\{t_0,\tau,Z\}$ 
obtained in case {\it a)}, by considering only the sample at $z\sim 0.3$,
and in case {\it b)}, by considering the properties of both the distant
sample and a nearby rich cluster, as discussed in the
Section.~\ref{FITMOD}.

In Figure~\ref{CPLOT1} we show the frequency distributions relative to
case {\it a)} in all the planes that can be constructed from the quantities
$ \{t_0,\tau,Z,K \}$, where both $K$ and $t_0$ refer to $z\sim0.3$.
Figure~\ref{CPLOT1} clearly shows the well known age--metallicity
degeneracy for which the photometric properties of older (younger)
stellar populations are equivalent to those of the more metal rich
(poor) ones.  This is particularly evident in the upper middle and
lower left panels as indicated by the elongation of the contours and
from the fact that very extended regions of the parameter space are
populated.

The most remarkable feature that arises from the comparison of the
cases {\it a)} and {\it b)} is the segregation in the space of
parameters obtained by constraining AC\,118 to belong to the same
evolutionary sequence of the local cluster.  Figure~\ref{CPLOT2} shows
that a large fraction ($\sim70\%$) of the points are constrained to
the region $ 0<\tau<3~\mathrm{Gyr}$, $5 <t_0< 9.0~\mathrm{Gyr}$ and
$0.008<Z<0.03$. The constraints at $z\sim0$ produce a sharp
metallicity sequence in the plane ($K$,$Z$), constraining brighter
galaxies to have higher values of $Z$. It is also interesting to
notice that about $20\%$ of the templates are not constrained to
follow a tight luminosity--metallicity relation, but are described by
higher values of the metallicity. A deeper inspection of
Figure~\ref{CPLOT2} shows that these objects are mostly found in the
region $\tau>3~\mathrm{Gyr}$ and $ t_0 > 4~\mathrm{Gyr}$ (formation
redshift $z_0>0.9$). Moreover, they do not show any significant
difference in their photometric properties with respect to the other
points of the model.

\subsection{Origins of the Color-Magnitude Relation}

KA97 and Kodama et al. (1998), by means of a population synthesis code
that accounts for chemical evolution in a self-consistent manner
(Arimoto \& Yoshii 1987), proved that the small evolution of the
CMR with look-back time constrains this relation to be a
metallicity-luminosity sequence.
In Figure~\ref{ZETASEQ} (upper panel) we compare the relation between
the luminosity-weighted mean stellar metallicity and the absolute
V-band magnitude at $z=0$ given by KA97 (see their Table~2), with
the same relation for our models. 
The points of the models were binned in the plane
($M_V$,$Z/Z_{\odot}$) with respect to V-band magnitudes and the
biweight estimator (e.g. Beers, Flynn \& Gebhardt 1990) was applied to 
derive the location of the
peak of the metallicity distribution at a given magnitude. Absolute
magnitudes were computed by a distance modulus for the Coma cluster of
$34.6~\mathrm{mag}$ (see KA97).
It is evident that the observed trend is fully consistent with the
findings of KA97. By using a least squares analysis, we find:
\begin{equation}
\log (Z / Z_{\odot}) = (-0.097 \pm 0.005) \cdot M_V + (-2.09 \pm 0.09).
\label{EQZSEQ}
\end{equation}
In order to investigate possible variations with luminosity of the age
of galaxies, we derive the relation between magnitude and formation
redshift $z_0$ for the objects that lie within $3\sigma$ of the
metallicity--luminosity relation. 
This distribution is shown in Figure~\ref{ZETASEQ}, bottom panel. The
formation epoch does not change along the sequence, and is constrained
to be greater than $z=1$ at the confidence level of $90\%$.  

\subsection{The Scatter of the Color Magnitude Relation}

So far, we have not yet discussed the constraints set by the present
analysis on the origin of the dispersion in the CMR. To this aim, we
computed for each galaxy the age $t_\alpha$ at which a given fraction
$\alpha$ of its stellar mass formed.  The parameter $t_{\alpha}$ is
given by the following combination of $t_0$ and $\tau$:
\begin{equation}
 t_{\alpha} = t_0 - \tau \cdot \ln(1-\alpha).
\label{EQTBURN}
\end{equation}
In Figure~\ref{TBURN} we plot the mean value of $t_{\alpha}$ (expressed
as redshift $z_{\alpha}$) as a function of the K-band magnitude at
$z=0.31$, and the relative percentiles of $16\%$ and $84\%$
(corresponding to a $1\sigma$ interval for a normal deviate). We chose
$t_{90\%}$, that corresponds to the age at which galaxies formed
$90\%$ of their stellar mass, and included only the points within the
%metallicity sequence. For $K \widetilde{<} 16.8$, the value of
metallicity sequence. For $\mathrm{K \lesssim 16.8}$, the value of
$z_\alpha$ is greater than $\sim1$ for almost all the points in the
model, while it decreases progressively at fainter magnitude. At
$\mathrm{K\sim18}$ ($\sim K^{\star}+3$) the redshift at which some
galaxies ceased to form most of their stars can be as low as
$z\sim0.5$.  On the contrary, the templates that lie outside the
metallicity sequence have larger values of $\tau$, and therefore
describe objects with a more recent star forming activity. By applying
Eq.~\ref{EQTBURN}, we find that all these objects did not complete to
form their stars at $z\sim0.3$.

\subsection{Global Star Formation History}

Finally we consider the global formation history of the stellar
populations in the galaxies of AC\,118. In Figure~\ref{SFRGLOB} we show,
as a function of redshift, the total (cumulative)
stellar mass $M(z)$ already formed at a given epoch in cluster galaxies.  
The function
$M(z)$ was obtained by summing the K-band luminosity-weighted mass
already formed at a given redshift $z$. Most of the luminous mass ($\sim50\%$)
present in the cluster at $z=0.31$ was formed at $z>2$ although star
formation continued at $z<1$ for $\sim20\%$ of the stars.

As noticed in previous studies (e.g. Poggianti et al. 1999, KB01), a
crucial role in estimating the star formation rate of cluster galaxies
can be played by the dust absorption. To investigate this subject, we
construct a simple model by assuming that all galaxies redder than the
red envelope of AC\,118 are obscured by a uniform screen of dust. We
use the differential dust extinction law introduced by Seaton (1979)
and adopt a color excess value $\mathrm{E(B\!-\!V)=0.1}$. Therefore, the
intrinsic magnitudes and colors of the templates are transformed
according to the equations $\Delta (\mathrm{R\!-\!K})=+0.44$ and 
$\Delta \mathrm{K=-0.09}$.  In Figure~\ref{dustps} we compare the 
CM diagram of
AC\,118 with the corresponding distribution of templates obtained by
including the dust effect in the model.  As can be seen, all the
galaxies of AC\,118 are properly represented by the model, including
the red outliers of the CM envelope.  These objects are described by
dusty blue spirals with extensive on-going star formation activity.
As a consequence, about $30\%$ (instead of $20\%$) of stars formed at
$z<1$ in this model, while the cumulative mass function decreases at
higher redshifts (see the dotted line in Figure~\ref{SFRGLOB}). We note
that, if the red outliers are actually blue dusty spirals, the adopted
value for the color excess corresponds to the minimum contribution of
the dust.

\section{Discussion and conclusions}\label{FINE}

We have studied the star formation history of galaxies in the rich
cluster AC\,118 at redshift $z=0.31$ by constraining their photometric
properties to reproduce, once evolved at $z\sim0$, those of a local 
rich cluster. The analysis is based on a large wavelength
baseline including accurate VRIK photometry for a large sample of
cluster galaxies (N=252). The sample was selected by the photometric
redshift technique and is complete in the NIR, thus reducing possible
biases towards objects with more recent/intense star formation
activity.

One of the main current issues in the comparison of the properties of
local and in\-ter\-me\-dia\-te--redshift clusters concerns the
selection criteria of the samples.  Studies of the CM relation based
on pure morphological selection can be biased towards the older
progenitors of nearby early-type galaxies (see van Dokkum et al. 2000 for
a detailed discussion).  On the other side, the application of a
statistical field subtraction approach requires a wide area around the
cluster field to be observed, while the use of a spectroscopically
selected sample at faint luminosities is very expensive in terms of
observing time (but not impossible, see Abraham et al. 1996; van
Dokkum et al. 2000).  The main advantage of a selection based on
photometric redshifts is that it allows to estimate the typical
luminosity-weighted formation epoch of a stellar population
irrespective of the past history of the host galaxy (such as, for
example, clustering through a merging hierarchy), and it is therefore
an ideal tool to define cluster membership for large samples of
galaxies without any tie to a particular scenario of galaxy formation.

A more tricky point is represented by the areas of the clusters to be
compared at different redshifts. In a hierarchical clustering picture,
clusters of galaxies are likely to accrete a significant fraction of
their population from the field even at relatively modest redshifts
($z<0.5$, see Kauffmann 1996). As a consequence, cluster richness
tends to increase with time, while the population accreted at an old
epoch becomes concentrated in a progressively smaller area (see
e.g. KB01).  On the other hand, the cores of rich
nearby clusters are very similar in their photometric properties: the
galaxy population is dominated by E/S0 galaxies with few Sp/Irr
having bluer colors. Moreover, early-type galaxies seem to follow
a universal well defined CM relation (see Bower et al. 1992a, Bower et al.
1992b).  For such reasons, we have analyzed the
constraints on the properties of the stellar populations of the
galaxies in AC\,118 by imposing that their evolution at $z \sim 0$
mimics the overall distribution in the ($\mathrm{K,V\!-\!K}$) plane for a
local rich cluster.

With the aim of constraining the galaxy evolution scenarios, several
studies have adopted a purely parametric approach, by comparing the
observed properties in the CM diagram with those predicted by models
that are based on different sets of parameters and that explore
different assumptions on the probability distributions of such
parameters.
These studies also assume that the scatter in the CM diagram arises
merely from age (but see Ferreras, Charlot, \& Silk 1999).  Although
the first epoch of star formation for the cluster early-type
population seems to be constrained to high redshifts for almost all
such models, further properties, as the last epoch and the spread of
star formation activity, are more model dependent.

The procedure we adopted describes each galaxy of AC\,118 by a stellar
population model, which is constrained, when evolved to $z\sim0$, to
be bounded by the red sequence locus or by the region of blue galaxies
of a rich nearby cluster (see Section 3.1). This is achieved by a
suitable procedure which scales the magnitudes of the galaxy
templates. We find that the best fitting models of AC\,118 galaxies
are able to match both the distributions in the ($\mathrm{K,R\!-\!K}$)
and ($\mathrm{R\!-\!K}$,$\mathrm{V\!-\!I}$) planes at $z\sim0.3$, and
the properties of the ($\mathrm{K,V\!-\!K}$) color-magnitude
distribution at $z \sim 0$, i.e. slope and intrinsic scatter of the CM
sequence, fraction of blue galaxies and luminosity function.  It is
important to notice that such a result is not implicit in the method
we used to scale the template magnitudes (see Section~\ref{FITMOD}).

\begin{figure}
\plotone{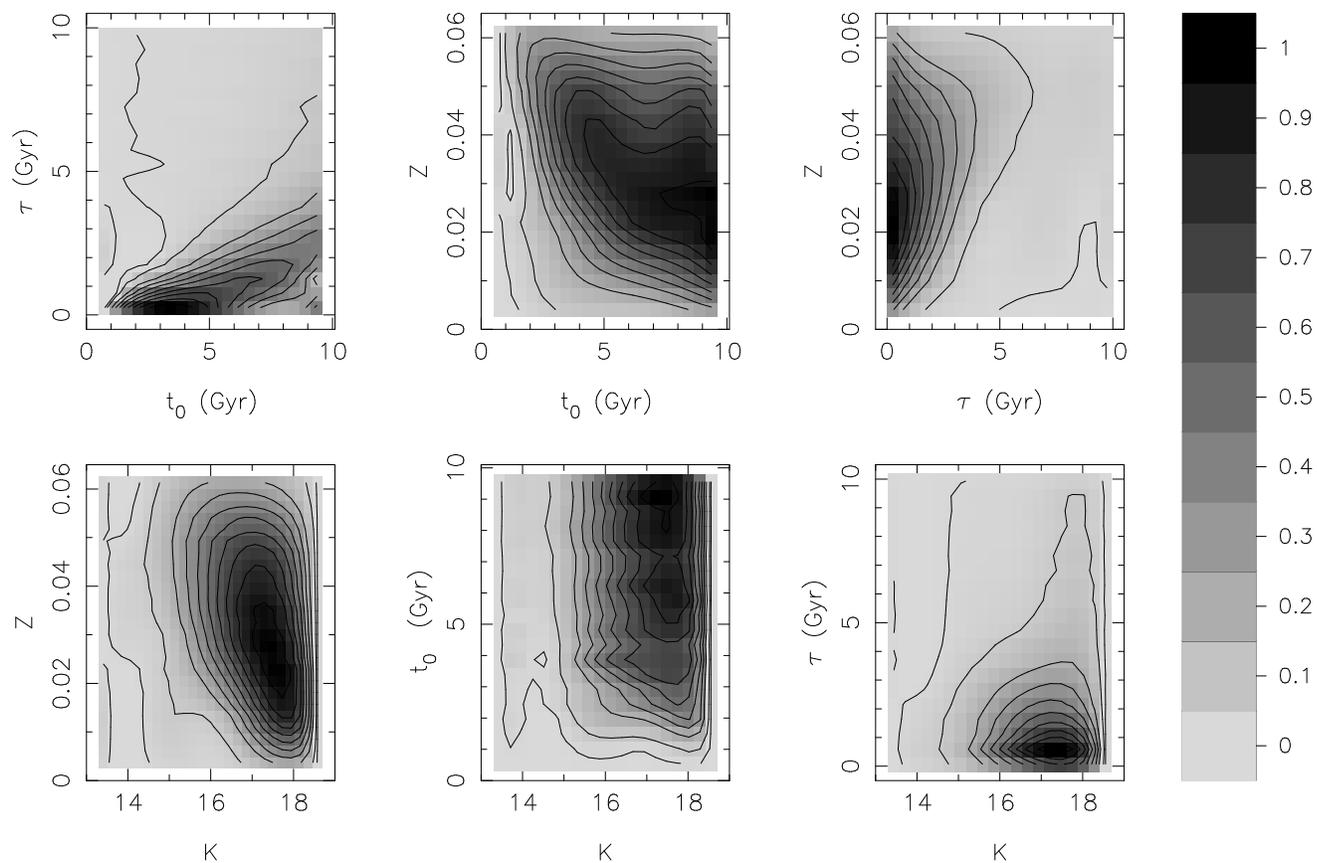}
\caption{Frequency distributions of the parameters that describe the
stellar populations in case {\it a)}, where the models are constrained 
to reproduce the color and magnitude distributions at $z\sim0.3$ only. 
Frequencies are normalized to unity (see the gray scale on the right).
The contours correspond to the frequencies  $0.01,0.07,0.2,0.3,0.4,0.8,0.9$.
}
\label{CPLOT1}
\end{figure}

\begin{figure}
\plotone{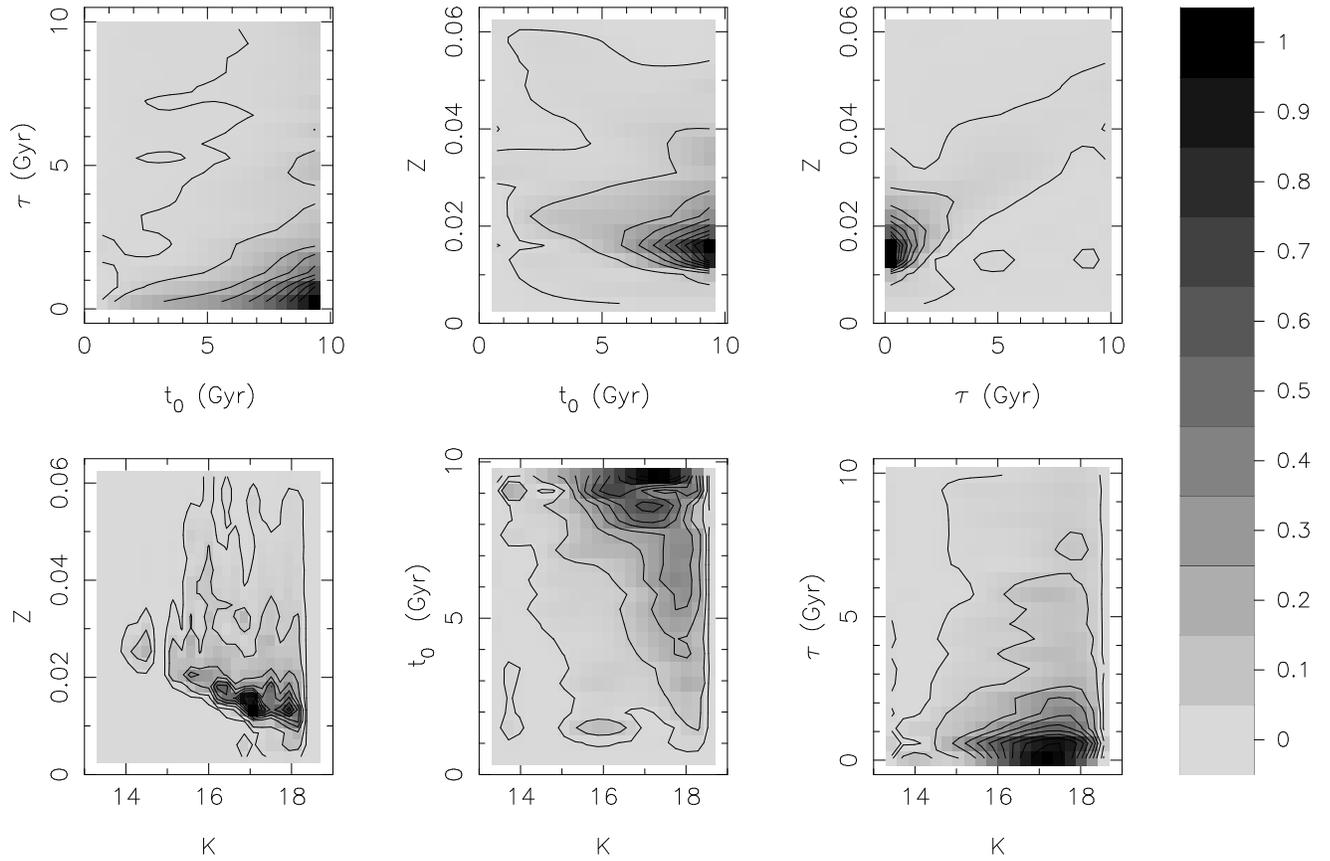}
\caption{The same as in Figure~\ref{CPLOT1} but for case {\it b)}, in
which the models are constrained by the properties of galaxies at both
$z\sim0.3$ and $z\sim0$. }
\label{CPLOT2}
\end{figure}

\begin{figure}
\epsscale{0.75}
\plotone{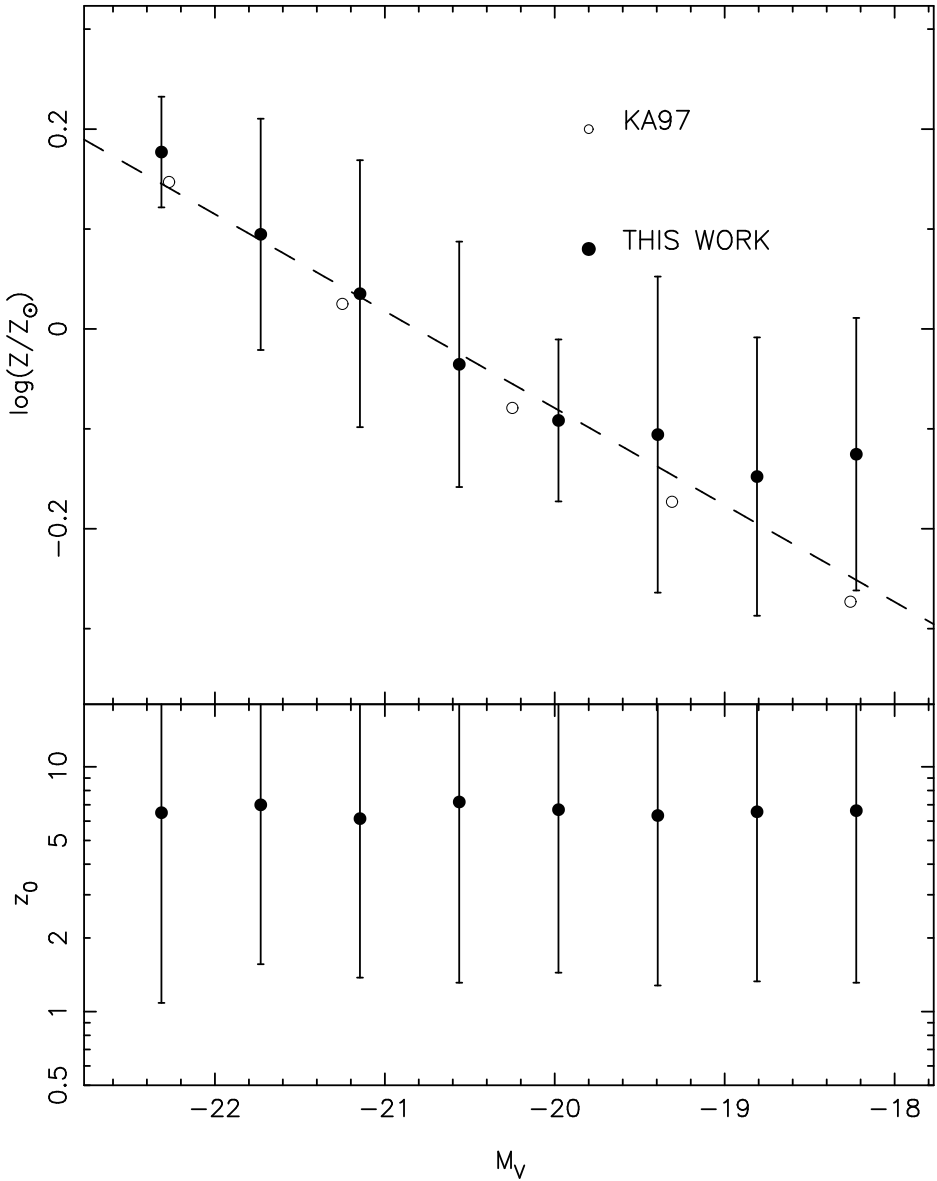}
\caption{ Metallicity $Z$ and formation redshift $z_0$ as a function
of absolute V-band magnitude at $z=0$. 
The bars connect the 16th and 84th percentiles of
the age and metallicity distributions at a given value of
$\mathrm{M_V}$. 
The dashed line in the upper panel corresponds to the 
line described by Eq.~\ref{EQZSEQ}. The open circles represent the 
luminosity-metallicity relation of Kodama \& Arimoto (1997).}
\label{ZETASEQ}
\end{figure}

\begin{figure}
\plotone{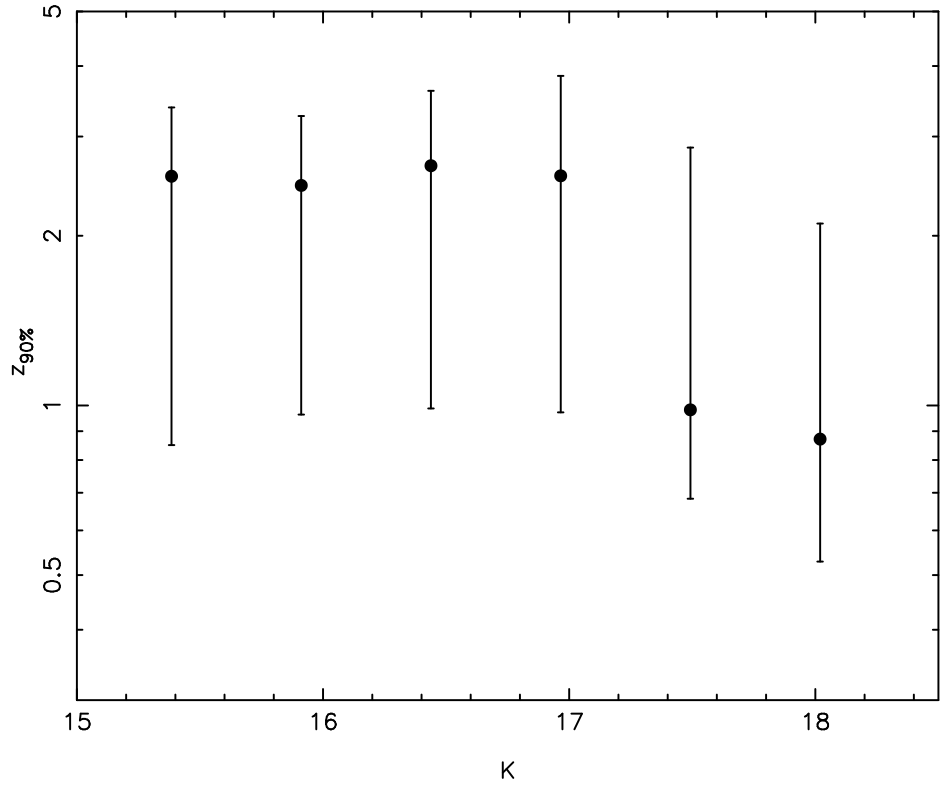}
\caption{Redshift $z_{90\%}$, at which 90\% of the stellar mass was formed,
vs. K-band magnitude at $\mathrm{z=0.3}$.
The bars connect the 16th and 84th percentiles.
}
\label{TBURN}
\end{figure}

\begin{figure}
\epsscale{1.05}
\plotone{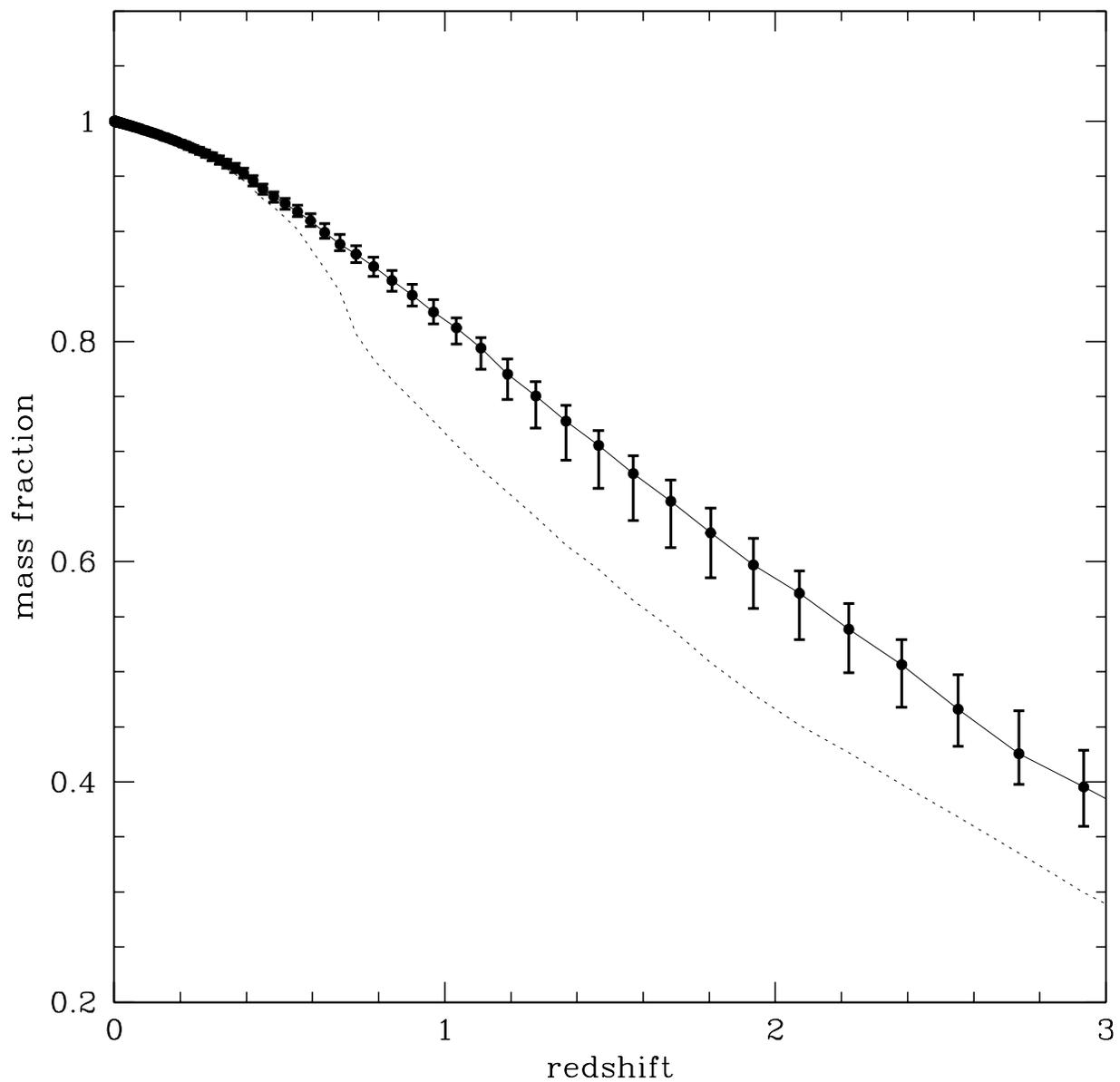}
\caption{Total mass of the stars at $z\sim0.3$ already formed in
cluster galaxies at a given redshift. The bars connect the 16th and
84th percentiles. The values of the redshift at which $M(z)$ are
calculated, are chosen in order to have a uniform spacing in look-back
time.  The dashed curves correspond to the model including dust
absorption.}
\label{SFRGLOB}
\end{figure}

\begin{figure}
\epsscale{1.05} 
\plotone{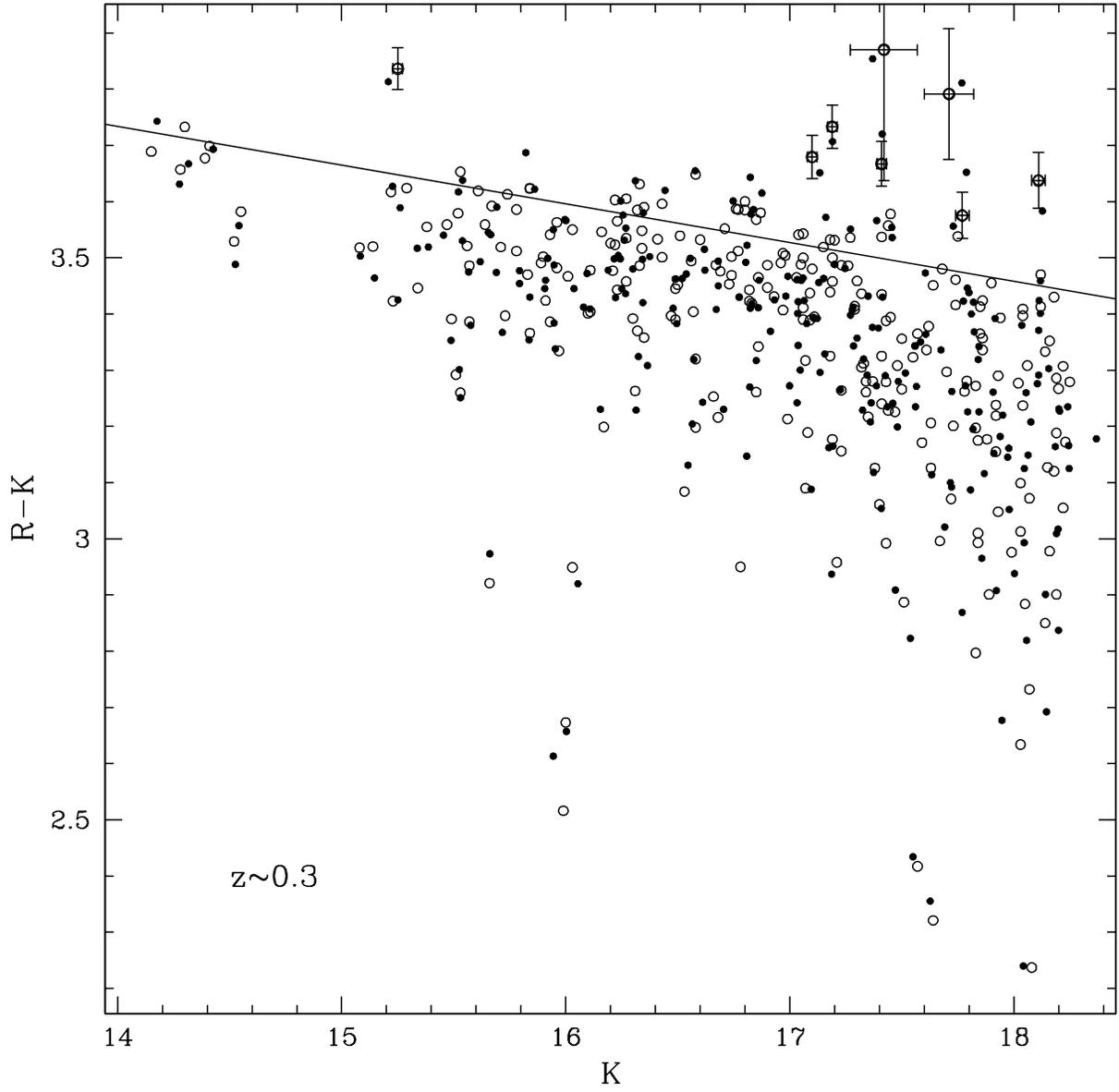}
\caption{CM diagram of galaxies in AC\,118 (open circles) and
of the model that include dust absorption (filled circles).
The distribution of the reddest galaxies is now recovered by the model 
(compare Figure~\ref{MODOBS}, upper left panel).}
\label{dustps}
\end{figure}

The constraint at $z \sim 0$ largely reduces the region of input
parameters available to the model. In particular, a sharp sequence
arises in the metallicity-luminosity diagram (cfr. lower left
panels of Figures~\ref{CPLOT1} and ~\ref{CPLOT2}), for which
brighter galaxies are described by higher values of $Z$. The
slope of the sequence is in full agreement with that derived by KA97 
in the framework of the monolithic collapse/galactic wind
model. It is interesting to notice that if we adopt the luminosity--weighted
mean stellar metallicity of the KA97 models, the zero-points of the relations
also coincide.  
The main difference between the results of KA97 and those of the
present work is that we do not obtain the metallicity sequence on the
basis of a particular galaxy evolution scenario.

The present data seem to exclude significant variations of star formation 
history along the CMR, and therefore confirm a pure metallicity interpretation,
in which the bulk of the populations formed
at high redshift ($z \sim 5$).  
These results, however, do not describe the properties of all the
stellar populations in AC\,118: we find that about $20\%$ of the
points of our model do not follow any metallicity--luminosity
relation, but are characterized by higher values of Z and more prolonged 
star formation activity ($\tau>4~\mathrm{Gyr}$).  
Since these objects do not show peculiar photometric properties in
the colors-magnitude space, this result could be the consequence of a
residual age--metallicity degeneracy. However, other possibilities can
be explored. For instance, the scatter of the CM relation at a
given luminosity could be partly due to the fact that more
recently assembled galaxies have higher metallicity than older systems of
similar luminosity (see Ferreras et al. 1999).

To study the dispersion in the CM diagram at $z\sim0.3$, we computed the
epoch $t_{90\%}$ at which galaxies completed to form $90\%$ of their
stars. While for $\mathrm{K<17}$ the corresponding redshift is greater than
$z=1$, at faintest magnitudes ($\mathrm{K}\sim K^{\star}+3$) we find that
some galaxies ceased to form stars at epochs as low as $z\sim0.5$.
These results are in agreement with the general picture that fainter
galaxies were more recently accreted from the field to the cluster
environment and therefore ceased to form stars at later epochs (see
KB01 for a wide discussion).

One half of the luminous mass present at $z\sim0.3$ formed at $z>2$, and
star formation continued at $z<1$ for $\sim 20\%$ of the stars. This
result changes if we are neglecting the effect of the dust obscuration
in a significant fraction of cluster galaxies. To investigate this
subject, we adopted a simple model in which all the galaxies redder
than the CM envelope at $z\sim0.3$ are obscured by a uniform screen of
dust.  The introduction of this model is also supported by the
presence of few galaxies of AC\,118 whose R-K color is too red with
respect to the CMR. These objects may be accounted for
as dusty galaxies with extensive on-going star formation
activity (cfr. KB01). In the model with dust, the
fraction of mass that forms at $z<1$ increases from $\sim 20 \%$ to
$\sim 30 \%$.

\acknowledgments

The observations at ESO were collected during the guaranteed time of 
the Osservatorio Astronomico di Capodimonte. 
Michele Massarotti is partly supported by a `MIUR-COFIN' grant.

\clearpage

%% Use the figure environment and \plotone or \plottwo to include 
%% figures and captions in your electronic submission.

%\begin{figure}
%\plotone{f1.eps}
%\caption{This is the first figure and it uses sgi9259.eps as
%its EPS figure file. \label{fig1}}
%\end{figure}

%% If you are not including electonic art with your submission, you may
%% mark up your captions using the \figcaption command. See the 
%% User Guide for details.
%%
%% No more than seven \figcaption commands are allowed per page, 
%% so if you have more than seven captions, insert a \clearpage 
%% after every seventh one. 

%% Tables should be submitted one per page, so put a \clearpage before
%% each one.

%% Two options are available to the author for producing tables:  the
%% deluxetable environment provided by the AASTeX package or the LaTeX
%% table environment.  Use of deluxetable is preferred.
%%

%% Three table samples follow, two marked up in the deluxetable environment,
%% one marked up as a LaTeX table.

%% In this first example, note that the \tabletypesize{}
%% command has been used to reduce the font size of the table.
%% Note also that the \label command needs to be placed 
%% inside the \tablecaption.

%\clearpage

\end{document}